\documentclass[%
        twocolumn,preprintnumbers,%
        prl,%
        floatfix,%
        aps,showpacs,amsmath,amssymb]{revtex4}
\usepackage[pdftex]{graphicx}

\newcommand{\tmpwidth}{}

\newcommand{\Ptorus}{P^{\rm T}}
\newcommand{\THeisenberg}{T_{\rm H}}
\newcommand{\Ttorus}{T_{\rm L}}
\newcommand{\epsC}{\epsilon_{\rm c}}

\newcommand{\ket}[1]{|{}#1{}\rangle}
\newcommand{\bracket}[2]{\langle{}#1{}|{}#2{}\rangle}

\begin{document}

\title{Dynamical Tunneling in Many-Dimensional Chaotic Systems}

\author{Akiyuki Ishikawa}
\author{Atushi Tanaka}%
\email{tanaka-atushi@tmu.ac.jp}
\author{Akira Shudo}
\affiliation{%
  Department of Physics, Tokyo Metropolitan University,
  Minami-Osawa, Hachioji, Tokyo 192-0397, Japan}

\date{\today} 

\begin{abstract}

We investigate dynamical tunneling 
in many dimensional systems using a quasi-periodically modulated
kicked rotor, 
and find that the tunneling rate from
the torus to the chaotic region is drastically enhanced when the chaotic
states become delocalized as a result of the Anderson transition.
This result strongly suggests that amphibious states, which were
discovered 
for a one-dimensional kicked rotor with transporting islands
[L. Hufnagel {\it et al.}, Phys. Rev. Lett. {\bf 89}, 154101 (2002)], 
quite commonly appear in many dimensional systems. 

\end{abstract}


\maketitle

According to the {\it semiclassical eigenfunction hypothesis},
each eigenfunction of a quantum system, whose classical counterpart
exhibits mixed-type phase space, is expected to be localized
on either regular or chaotic components of classical phase space
in the semiclassical limit~\cite{Percival73}.
In a finite $\hbar$ regime, however, the tunneling effect comes in
and it hybridizes localized eigenfunctions together.
This has invoked considerable interest in quantum tunneling
in two or more degrees of freedom systems~%
\cite{CATRAT,CreaghTunnel,Creagh98,Shudo95,Takahashi00}.

Since the tunneling effect is supposed to be exponentially small in  
general,
one may regard tunneling merely as a correction
to the semiclassical eigenfunction hypothesis.
However, it was found that under a certain condition eigenfunctions
are not necessarily localized on either regular or chaotic regions
even when $\hbar$ is much smaller than the area of the regular region.
Thus
it looks to be violating the  semiclassical
eigenfunction 
hypothesis~\cite{Hufnagel-PRL-89-154101,Baecker-PRL-94-054102}.

Such states are called {\it amphibious eigenstates} and their origin
was discussed in Ref.~\cite{Baecker-PRL-94-054102}:
They pointed out the importance of the relation between two time scales,
the Heisenberg time $\THeisenberg$ in the chaotic sea and the torus's life time 
$\Ttorus$,
to understand such exotic states~\cite{note:BaeckerInTermsOfEDOS}.
Since the former is associated with
the localization length in the chaotic sea, the argument on the two  
competing time scales
can be rephrased using the interplay between 
dynamical localization~\cite{Casati-LNP-93-334-short} and dynamical tunneling.

A more direct explanation for the ``flooding'' of the wavefunction~%
\cite{Hufnagel-PRL-89-154101,Baecker-PRL-94-054102}
can be given based on the time-domain semiclassical picture~%
\cite{Shudo09}.
In particular, the theory of complex dynamical systems tells us that  
there are
exponentially many complex orbits which connect the torus and chaotic  
regions.
It could further be proved that they (i) act as tunneling orbits when  
they stay in the torus region,
and (ii) behave as if they are real orbits after reaching the chaotic  
sea~\cite{Shudo09}.
In other words, dynamical tunneling and dynamical localization 
are controlled by the same orbits with amphibious characters,
i.e., both regular and chaotic ones.
Therefore, destructive interference among the semiclassical orbits  
causing
the suppression of classical diffusion~\cite{Shudo94} simultaneously  
inhibits
chaotic tunneling from the torus to the chaotic sea. On the contrary,
the attenuation of dynamical localization and the recovery
of the chaotic tunneling occur
simultaneously~\cite{Ishikawa-PRE-80-046204-debug}.

Amphibious states have been considered to appear in quite a specific 
situation where the system has 
accelerator modes~\cite{Hufnagel-PRL-89-154101,Baecker-PRL-94-054102}.
However, 
the accelerator modes are introduced just 
to prepare extremely large localization 
lengths~\cite{Baecker-PRL-94-054102}. 
The system with large localization lengths can also be realized 
in many degrees of freedom systems 
according to the correspondence 
of the dynamical localization to the Anderson 
localization~\cite{Fishman-PRL-49-509,Casati-PRL-62-345,Doron-PRL-60-867,%
 Chabe-PRL-101-255702-short}.
This naturally leads to a conjecture: {\em The appearance of amphibious 
states is a common feature 
in many-degrees of freedom systems.}
This, however, needs to be scrutinized, since  
B\"acker {\it et al.}'s argument in Ref.~\cite{Baecker-PRL-94-054102}
provides only a sufficient 
condition of the absence of 
amphibious states under the small localization
lengths,
and no condition that ensures the 
amphibious states is known.
In this Letter, we will provide 
numerical evidence of the conjecture.

\renewcommand{\tmpwidth}{6.83cm} 
\begin{figure}
 \centering
 \includegraphics[width=\tmpwidth]{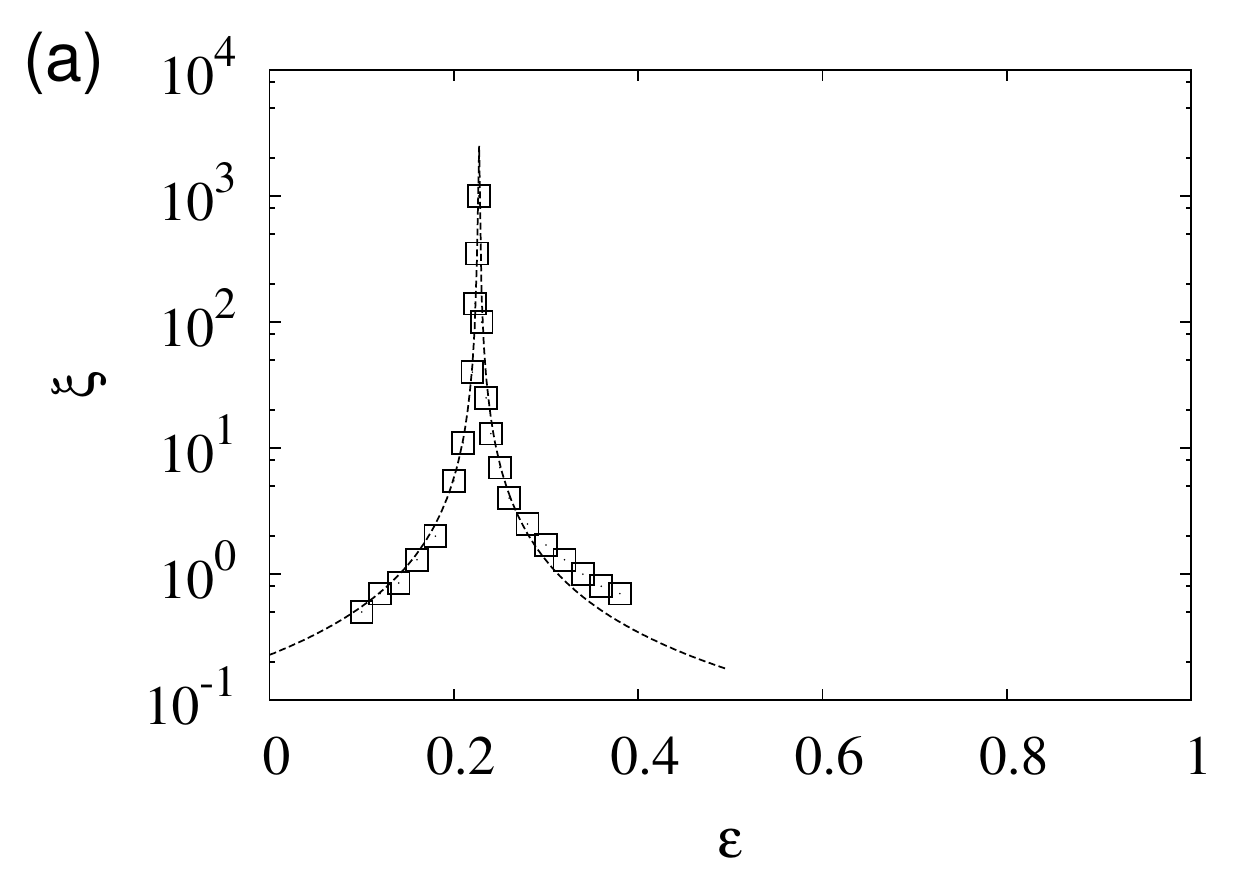}
 \\
 \includegraphics[width=\tmpwidth]{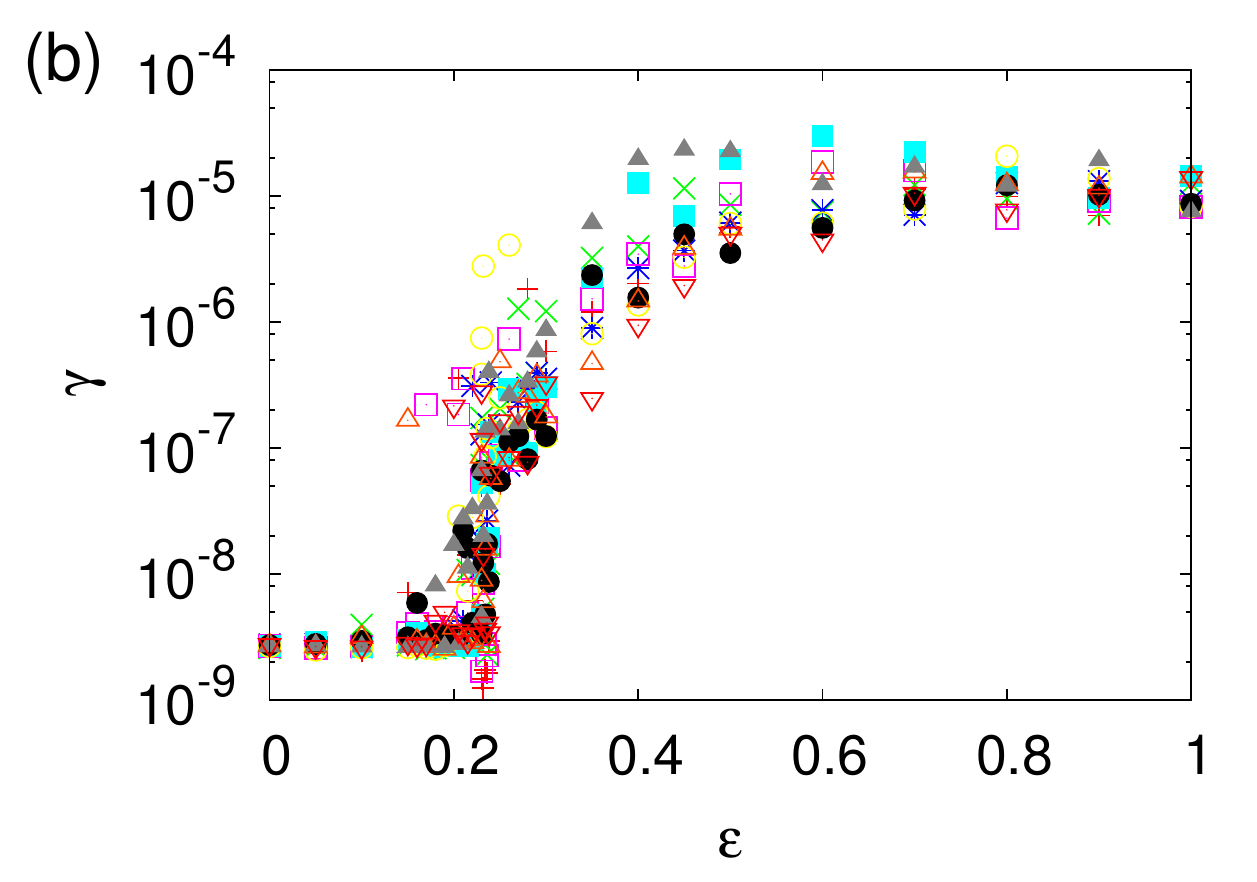}
 \caption{%
   (Color online)
   (a) The scaling parameter $\xi$, which is proportional to the 
   localization length and the inverse of the diffusion constant
   in the insulator and metallic phases, respectively, as a function of
   the strength of the modulation 
   $\epsilon$ in our model [Eq.~\eqref{eq:defHC}].
   $T(p)$ and $g(p)$ are chosen
   so as to make the whole phase space chaotic 
   (see the main text).
   $\xi$ diverges at the critical point 
   $\epsC\simeq0.227$.
   A thin curve indicates $\xi= a |\epsilon - \epsC|^{-\nu}$ 
   with $\nu=1.55$ and $a=2.4\times 10^{-2}$.
   The parameters are set to be
   $k=2$, $s=4 +\sqrt{3}/10$, $b=0$, $M=2$,
   $\omega_1 = 2\pi\sqrt{5}$, $\omega_2 = 2\pi\sqrt{13}$,
   $W_p = 1000$, and Planck's constant $h=0.2$.
   (b) 
   The decay rate $\gamma$ of $\Ptorus_n$ [Eq.~(\ref{eq:defPtorus})]
   as a function of $\epsilon$. 
   Different marks correspond to ten different values of 
   $s$ for a small interval $4.17 < s < 4.19$.
   For
   $\epsilon \ll \epsC$,
   chaotic tunneling is almost forbidden, i.e., 
   $\gamma\sim 0$~\cite{note:smallGamma}.
   Around $\epsilon\sim\epsC$, $\gamma$ exhibits strong
   fluctuation, which becomes smaller for larger $\epsilon$.
   The parameters for Eq.~(\ref{eq:defH}) are $b=5.6-W_p/2$, 
   $\tilde{b}= b + 0.2$,
   $\omega=\sqrt{3}$ and $\beta=50$.
   The initial state $\ket{\psi_0}$ satisfies the EBK condition
   whose center of the momentum is at $p=b - 0.6$ 
   (see, Appendix of Ref.~\cite{Ishikawa-PRE-80-046204-debug}).
 }
 \label{fig:eps-gamma}
\end{figure}

We examine a one degree of freedom kicked rotor whose phase space $(q,p)$ 
is divided into a torus region $p \lesssim b$ and a chaotic sea 
$p \gtrsim b$, which are connected by dynamical tunneling.
To introduce many-dimensionality in effect, 
the system is described by the kicked Hamiltonian with 
quasi-periodic modulations~\cite{Casati-PRL-62-345,Yamada-PLA-328-170}:
\begin{align}
 \label{eq:defHC}
 H 
 &= 
   T(p)
 + V(q)\sum_{n}\delta(t-n)
 \nonumber\\&\qquad
 + \epsilon 
   g(p)
 \frac{1}{M}\sum_{j=1}^M \cos(\omega_j t)
 \sum_{n}\delta(t-n-0)
 ,
\end{align}
where $V(q) = k\cos(2\pi q)/(4\pi^2)$,
and modulation frequencies $\{\omega_j/(2\pi)\}_{j=1}^M$ are 
irrational and non-resonant with each other~\cite{note:Floquet}.
$T(p)$ and $g(p)$ are specified in the following.
We impose periodic boundary conditions
$0\le q < 1$ and $-W_p/2 \le p < W_p/2$~\cite{note:diffraction}.

Before studying the mixed case, we first examine the fully chaotic case 
to identify insulator and metallic phases in the parameter
space of $\epsilon$, with a sufficiently large value of $W_p$.
We choose  
$T(p) = s (p-b)^2/2$ and $g(p) = p - b$.
In the absence of modulation, i.e, $\epsilon=0$, the mapping 
derived
by Eq.~(\ref{eq:defHC}) can be reduced to the standard mapping whose 
nonlinearity parameter is $K=sk$. In the following, we examine 
the case that the nonlinearity is sufficiently large $K \sim 8$,
where the phase-space is mostly filled with 
a chaotic sea~\cite{LichtenbergLieberman-1992}.
When $\epsilon$ is non-zero, this model is essentially the same as 
Casati {\it et al.}'s quasi-periodic kicked rotor,
which is equivalent with a $M+1$ dimensional tight-binding model 
with pseudo-disorder and is shown to exhibit the Anderson
transition with $M=2$~\cite{Casati-PRL-62-345,Chabe-PRL-101-255702-short}.
We also examine the case $M=2$ 
and numerically confirmed that our model is in 
insulating and metallic phases at $\epsilon < \epsC$ and
$\epsilon > \epsC$, respectively, where 
$\epsC \simeq 0.227$ is the metal-insulator transition 
point.
Using the finite size scaling technique~\cite{%
 Ohtsuki-JPSJ-66-314,Slevin-PRL-82-382,Chabe-PRL-101-255702-short},
we obtained numerical evidence of the Anderson transition
in the chaotic sea. 
The details will be reported elsewhere~\cite{IshikawaFull}.
We show how the scaling parameter $\xi$~\cite{Slevin-PRL-82-382}, 
which is proportional to the localization length and the inverse 
of diffusion constant in the insulating and metallic phase, respectively,
depends on $\epsilon$ in Fig.~\ref{fig:eps-gamma}(a).

\begin{figure}
 \centering
 \includegraphics[width=6.23cm]{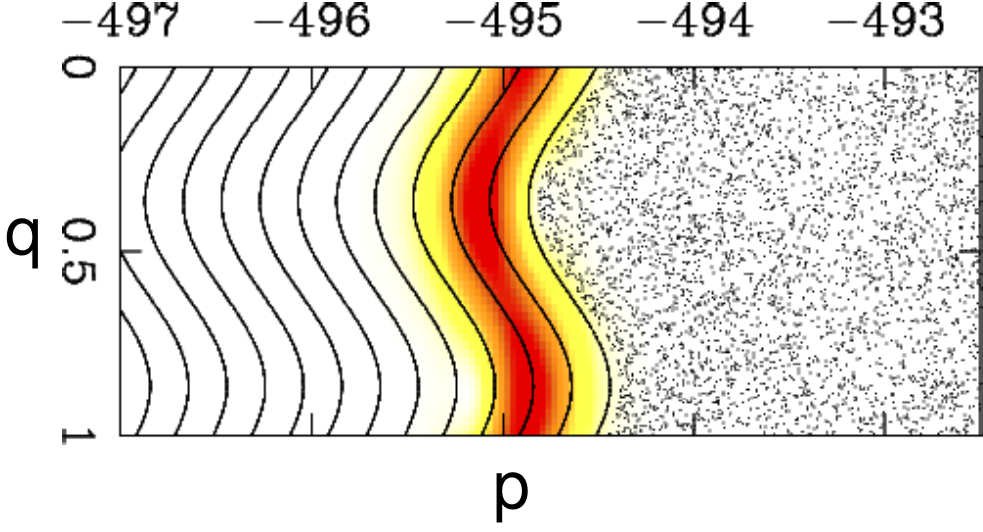} 
 \caption{%
   (Color online)
   Poincar\'e section around the torus region of the tunneling degrees of 
   freedom under the absence of modulation (i.e., $\epsilon=0$).
   The torus region $p\lesssim b$ and the chaotic sea $p \gtrsim b$ are 
   seen to be sharply divided.
   The density plot of the Husimi function 
   of the initial torus state, 
   which is constructed by EBK quantization, is superposed.
   The parameters are same as in Fig.~\ref{fig:eps-gamma}(b), except
   $s=4 +\sqrt{3}/10$.
 }
 \label{fig:model}
\end{figure}

Next,
we introduce a torus region in $p \lesssim b$ with keeping 
untouched the chaotic nature of the region $p \gtrsim b$ 
(see, Fig.~\ref{fig:model}),
using the following kinetic term~\cite{Ishikawa-PRE-80-046204-debug}
\begin{align}
 \label{eq:defH}
 T(p)
 &\equiv
 \frac{s}{2}(p-b)^2 \theta_{\beta}(p - b) + \omega(p-b)
 ,
\end{align}
where 
$\theta_{\beta}(x) = [1 +\tanh(\beta x)]/2$ 
is a smoothed step function with 
a smoothing parameter $\beta$~\cite{note:quantumAnomaly}.
The region $p \lesssim b$ is filled with tori because
$T(p)$ is effectively linear there.
At the same time, 
we employ the modulation term
$g(p)\equiv (p - \tilde{b})\theta_{\beta}(p - \tilde{b})$,
where $\tilde{b}$ is slightly larger than $b$.
This makes the torus region almost independent of the modulation.
Hence, only dynamical tunneling induces transitions between 
the torus region and the chaotic sea.

\begin{figure}
 \centering
 \includegraphics[width=8.6cm]{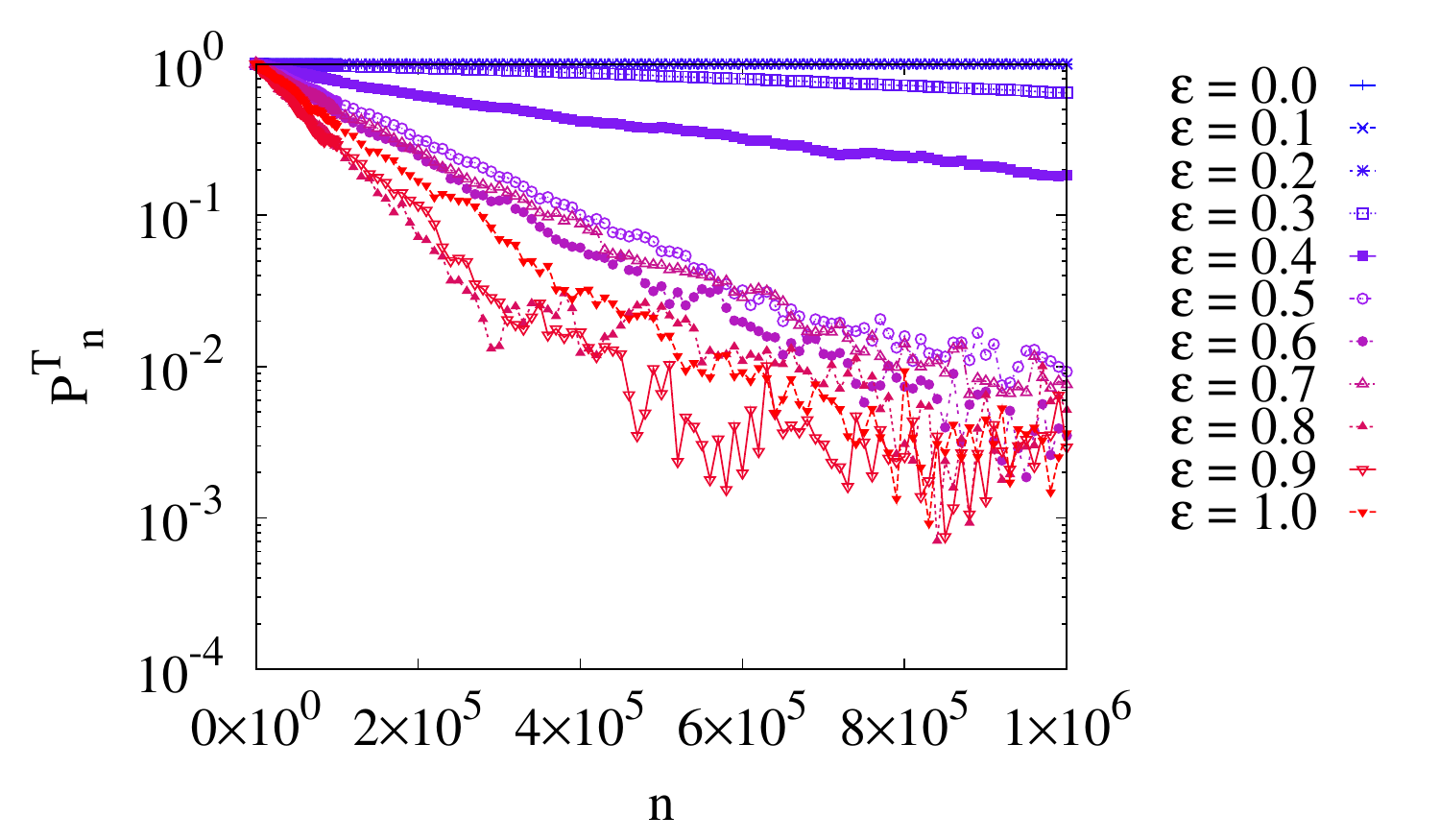}
 \caption{(Color online)
   The survival probability $\Ptorus_n$ of a torus state
   as a function of the number of kicks $n$.
   The
   parameters are the same as in
   Fig.~\ref{fig:model}.
   When the chaotic sea is in the insulating phase 
   $\epsilon < \epsC$,
   the tunneling decay is almost forbidden.
   For $\epsilon \ge 0.3$ ($>\epsC$), $\Ptorus_n$
   exhibits exponential decay.
   Around the transition point, the decay of $\Ptorus_n$
   strongly depends on $\epsilon$, and 
   the decay is too slow to determine whether or not
   $\Ptorus_n$ obeys the exponential law.
 }
 \label{fig:Ptorus}
\end{figure}

Tunneling leakage from the torus to chaotic regions is monitored by
the integration of $|\bracket{p}{\psi_n}|^2$ for the whole
torus region~\cite{note:PtorusRange}
\begin{align}
 \label{eq:defPtorus}
 \Ptorus_n \equiv \int_{-W_p/2}^{b} |\bracket{p}{\psi_n}|^2 dp
 ,
\end{align}
where $\ket{\psi_n}$ denotes the state vector 
at time step $n$.
We prepare the system initially to be in a torus state $\ket{\psi_0}$, 
which satisfies the Einstein-Brillouin-Keller (EBK) quantization 
in the torus region (see, Fig.~\ref{fig:model}).
Time evolution of $\Ptorus_n$ strongly depends on $\epsilon$, as shown
in Fig.~\ref{fig:Ptorus}. When $\epsilon$ is far below 
$\epsC$, $\Ptorus_n$ keeps almost unity
within, say, $10^{6}$ steps.
As shown in Fig.~\ref{fig:wavefunc}(a), 
the corresponding wave function in the momentum representation exhibits 
dynamical localization in the chaotic sea.
As was shown in \cite{Ishikawa-JPA-40-F397}, and will be reported in 
\cite{IshikawaFull}, there appear exponentially many complex orbits
connecting the torus and chaotic region, nevertheless such ``flooding'' of
complex orbits are suppressed because of the destructive interference 
in the chaotic region.

For larger $\epsilon$, chaotic tunneling recovers 
(see, Fig.~\ref{fig:Ptorus}).
Figure~\ref{fig:eps-gamma}(b) shows 
the $\epsilon$-dependence of decay rate $\gamma$,
which is obtained by fitting of $\Ptorus_n$.
Around $\epsC$, 
$\gamma$ changes significantly
with large fluctuations.
A strong correlation between $\epsilon$ dependences of
$\gamma$ and $\xi$ is evident.
Indeed, when $\gamma$ is significantly different from zero, 
the classical diffusion in the chaotic sea $p\gtrsim b$ is recovered,
as shown in Fig.~\ref{fig:wavefunc}(b).
This strongly suggests that chaotic tunneling recovers 
when the chaotic sea is in the metallic phase. 
Note that the localized component on the torus region almost 
disappears at $n = 10^6$, and the tails of wavefunction is 
clearly given as Gaussian 
(see inset of Fig.~\ref{fig:wavefunc}(b)).

We remark on the fluctuation of $\gamma$ (see, Fig.~\ref{fig:eps-gamma}(b)).
Around $\epsC$, $\gamma$ strongly depends on
$s$ as well as $\epsilon$.
A possible origin of the fluctuation is the resonance induced
by near degeneracies among approximate quasienergies of torus and 
chaotic states~\cite{Ishikawa-PRE-80-046204-debug}. 
Such a strong fluctuation 
against
parameter variations
is a common feature of dynamical tunneling in nonintegrable systems~%
\cite{CATRAT}.
For smaller $\epsilon$ ($\ll \epsC$), the resonances
are ineffective~\cite{note:resonance}.
In contrast, the effect of the resonances becomes prominent around $\epsC$.
This is because 
the effective density of states around 
the torus state become larger due to the exponential growth
of the localization length in the chaotic sea 
(see, Fig.~\ref{fig:eps-gamma}(a)).
For much larger $\epsilon$ ($\gg\epsC$), the fluctuation becomes
smaller. 
This is supposed to be due to
the completion of the transition to the metallic phase. 
 In other words, the effective density of states around 
 the torus state become so large, the fluctuations induced by
 the resonances are averaged out.

\renewcommand{\tmpwidth}{7.0cm}
\begin{figure}
 \centering
 \includegraphics[width=\tmpwidth]{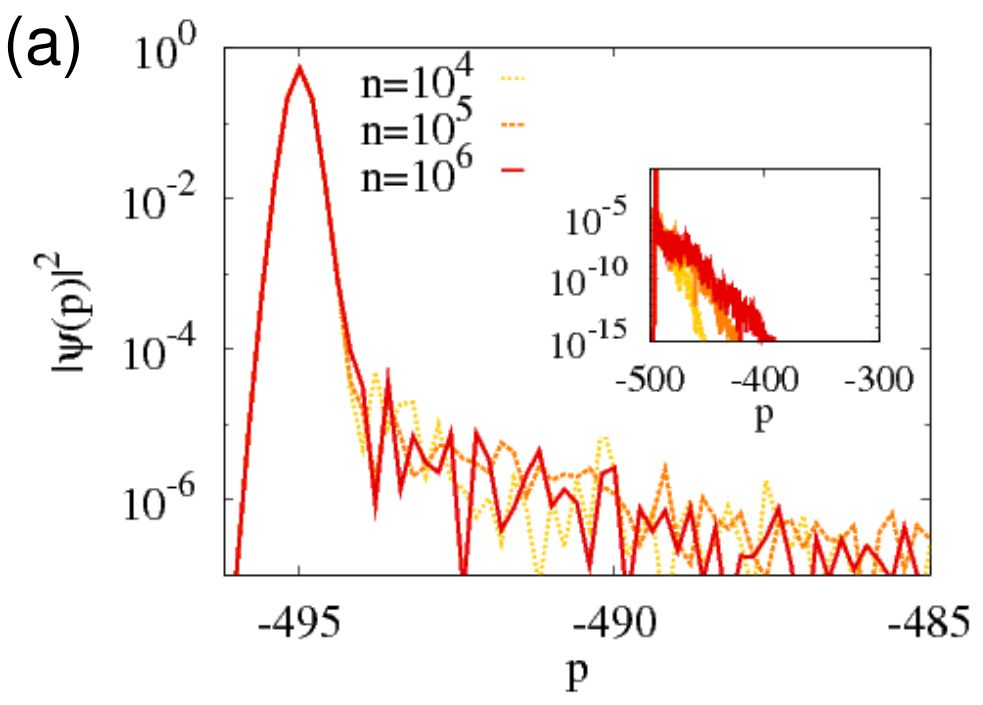}
 \hfill
 \includegraphics[width=\tmpwidth]{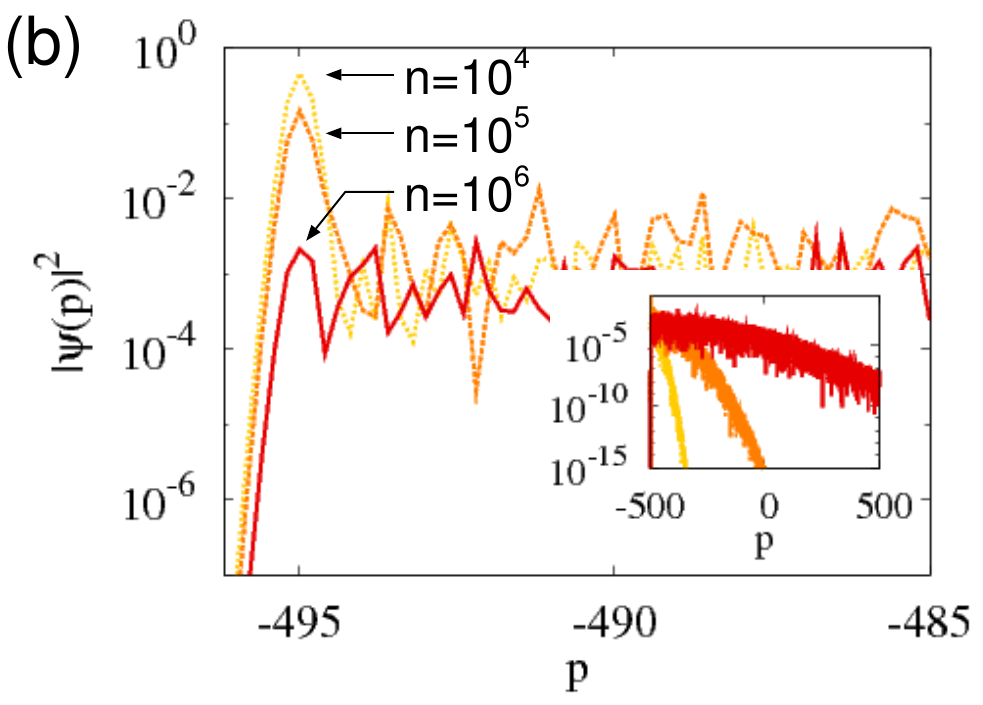}
 \caption{%
   (Color online)
   Snapshots of the momentum distribution $|\bracket{p}{\psi_n}|^2$.
   (a) In the insulating phase ($\epsilon=0.2$),
   $|\bracket{p}{\psi_n}|^2$
   remains almost unchanged in the torus region and 
   dynamical localization occurs in the chaotic sea. 
   (b) In the metallic phase ($\epsilon=0.8$), 
   $|\bracket{p}{\psi_n}|^2$
   decays in the torus region and the envelope of
   $|\bracket{p}{\psi_n}|^2$
   in the chaotic sea around the torus region
   is almost unchanged.
   However, as is seen in the inset of (b), the diffusion toward the uniform 
   distribution in the momentum space is not completed for $n<10^6$.
 }
 \label{fig:wavefunc}
\end{figure}

So far, we have
focused on the case that $W_p$ is sufficiently large.
As for the ``phase transition''
between the suppression and restoration of chaotic tunneling,
our numerical result suggests that the convergence for 
the ``thermodynamic'' limit $W_p\to\infty$ is quite fast.
In this idealized limit, one may regard that the delocalized 
chaotic sea plays a role of ``particle bath''.

To be precise, however,
the transition from the suppression of chaotic tunneling
to the recovery is 
not sharp, but rather smooth,
as seen in Fig.~\ref{fig:eps-gamma}(b).
This is because the recovery occurs even in the insulating 
phase with sufficiently 
large localization length due to the B\"acker 
{\it et al.}'s condition 
$\Ttorus \simeq \THeisenberg$~%
\cite{Baecker-PRL-94-054102},
which determines the border between the suppression and the recovery.

\begin{figure}
 \centering
  \includegraphics[width=7.00cm]{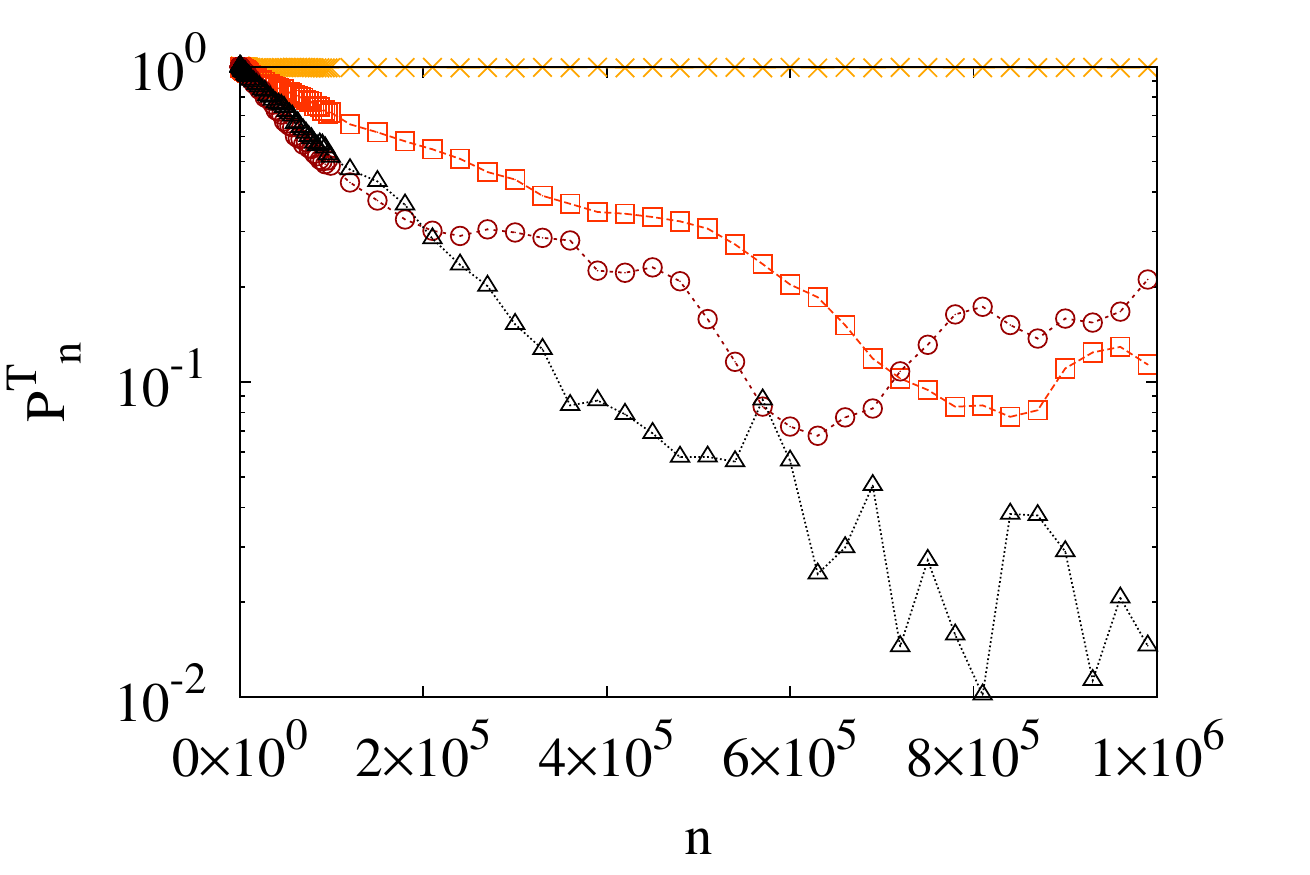}
  \caption{(Color online)
    Evolution of $\Ptorus_n$ for the case of smaller $W_p = 30$.
    We choose $\epsilon=0.5$ ($>\epsC$).
    To ensure the spectrum of the system in the delocalized ``phase''
    to be discrete one, the modulation frequencies
    are chosen to be rational~\cite{Casati-PRL-62-345},
    i.e., $\omega_1/(2\pi) = 682/305$ and 
    $\omega_2/(2\pi) = 11/3$ ($\times$), 
    $119/33$ ($\square$), 
    $649/180$ ($\bigcirc$),
    and $4287/1189$ ($\triangle$).
    Other parameters are the same as in Fig~\ref{fig:Ptorus}.
  }
 \label{fig:SmallWp}
\end{figure}

Furthermore, the crossover between the suppression and the restoration of 
chaotic tunneling occurs even when
$W_p$ is rather small.
If $\epsilon$ is sufficiently larger than $\epsilon_{\rm c}$,
the time evolution of $\Ptorus_n$ obeys the irreversible decay
for a short time period, and after that, exhibits 
erratic oscillations (see, Fig.~\ref{fig:SmallWp}).
Although the physical picture based on
the thermodynamic limit is inapplicable anymore,
our numerical result indicates 
the presence of a considerable number of amphibious states, 
i.e., eigenstates that have significant 
overlap with both torus and chaotic regions.

We have numerically investigated the stability of torus states surrounded 
by chaotic seas in many degrees of freedom systems. 
If the coupling strength exceeds a critical value at which 
the Anderson transition occurs in the chaotic region, the nature of 
dynamical tunneling drastically changes. 
The result of the smaller $W_p$ case 
may give rise to a reexamination of 
quantization condition of many dimensional mixed systems. 
Although the quantization of tori and chaotic seas
can be carried out separately in the semiclassical limit, 
this seems not to be the case even when 
the size of Planck's constant is considerably smaller
than the area of a torus region, due to the emergence of
amphibious states.
The ``nonseparability'' among regular and chaotic region certainly
has been a problem of tunneling in nonintegrable 
systems. Our result suggests that the occurrence of the nonseparability,
i.e., the emergence of amphibious states, 
is the rule rather than the exception in many-dimensional mixed systems.

\begin{acknowledgments}
We wish to thank Hiroaki Yamada and Kensuke Ikeda for discussion.
This work has been supported by the Grant-in-Aid for Scientific Research of
MEXT, Japan (Grant number {21540394}).
\end{acknowledgments}


\begin{thebibliography}{29}
\bibitem[{not({\natexlab{a}})}]{note:BaeckerInTermsOfEDOS}
\bibinfo{note}{The B{\"a}cker {\it et al.}'s
  condition~\cite{Baecker-PRL-94-054102} can also be understood as the
  competition between the mean level spacing in the chaotic sea, and the energy
  width of the torus state in the limit that the torus state become unstable.}
\bibitem[{not({\natexlab{b}})}]{note:smallGamma}
\bibinfo{note}{Since we determine $\gamma$ from a finite time series of
  $\Ptorus_n$ ($n < 10^6$), smaller $\gamma$ (say, less than $10^{-7}$) should
  not be interpreted as a decay rate (cf. Fig.~\ref{fig:Ptorus}).}
\bibitem[{not({\natexlab{c}})}]{note:Floquet}
\bibinfo{note}{From Eq.~\eqref{eq:defHC}, 
  the Floquet operator at the $n$-th step is 
  $ \hat{U}_n = e^{-iT(p)/\hbar}
  e^{-i\epsilon g(p)\sum_{j=1}^M \cos(\omega_j n)/(M\hbar)} 
  e^{-iV(q)/\hbar}$.}
\bibitem[{not({\natexlab{d}})}]{note:diffraction}
\bibinfo{note}{We numerically confirmed that the diffraction due to the
  discontinuity of the kinetic term at $p=\pm W_p/2$ is negligible here.}
\bibitem[{not({\natexlab{f}})}]{note:PtorusRange}
\bibinfo{note}{Although $\Ptorus$ can contain a contribution from the chaotic
  sea, such a contribution immediately diffuse out. Hence it is negligible on
  average.}
\bibitem[{not({\natexlab{g}})}]{note:resonance}
\bibinfo{note}{In Ref.~\cite{Ishikawa-PRE-80-046204-debug}, the resonances at
  the absence of noise are more prominent than the present result, because the
  initial torus state is chosen to be nearer to the chaotic sea.}
\end{thebibliography}


\end{document}